# The highest-energy cosmic-rays – the past, the present and the future

*Alan* Watson[*]

School of Physics and Astronomy, University of Leeds, Leeds LS2 9JT, UK

**Abstract.** The greater part of this paper is concerned with a historical discussion of the development of the search for the origins of the highest-energy cosmic-rays together with a few remarks about future prospects.

Additionally, in section 6, the situation with regard to the mass composition and energy spectrum at the highest energies is discussed. It is shown that the change of the depth of shower maximum with energy above 1 EeV, measured using the Telescope Array, is in striking agreement with similar results from the Auger Observatory. This implies that either the mean mass of cosmic rays is becoming heavier above ~4 EeV or that there is a change in details of the hadronic interactions in a manner such that protons masquerade as heavier nuclei. A long-standing controversy is thus resolved: the belief that pure protons dominate the mass distribution at the highest energies is no longer tenable.

## 1 Introduction

In this paper I try to trace the key steps taken to discover the origins of the highest-energy particles in Nature. Advances in science are always driven by advances in techniques and ideas for what should be measured. What I find remarkable is that the present flourishing of activity on the highest-energy cosmic-rays relies almost entirely on concepts, both analytical and technical, that were proposed more than 40 years ago and by a handful of people. V L Fitch has pointed out that those interested in cosmic rays tended, in earlier times, to be rugged individualists, to be iconoclasts, and to march to the drummer in their own heads rather than some distant one [1]. Today the problems and solutions identified by the ancients are tackled by the large collaborations needed to command the funding and to operate the large and complex instruments required to progress the field. Below I have tried to describe some of these early efforts and to highlight the contributions of those individuals – the rugged iconoclasts – who lit our way. Growth of computing has also been a boon though much was done and understood about shower development before the mid-1960s when shower simulations came to have a major impact.

I conclude with some remarks on the present status of work on the mass composition and energy spectrum at the highest energies, in the case of the former shedding light on a long-standing puzzle. I also make a few remarks about future prospects.

## 2 The Early Years

The story of the discovery of cosmic rays by Victor Hess in 1912 needs no re-telling. He certainly fits Fitch's description - with the addition of extraordinary bravery. What is perhaps less well-known is that it was a Scottish physicist, C T R Wilson, who in 1901 was the first to suggest that the problem of the discharge of well-insulated charged objects – a problem that challenged some of the best physicists of the late nineteenth century – might be explained by invoking a non-terrestrial agency. Wilson wrote: *'the continuous production of ions in dust-free air could be explained as being due to radiation from sources outside our atmosphere, possibly radiation like Röntgen rays or cathode rays but of enormously greater penetrating power.'* [2] To test his speculation he took electroscopes into a railway tunnel south of Edinburgh, expecting to see the rate of discharge greatly slowed because of the shielding of the rocky over-burden. At this time radioactivity in rocks was little understood and the anticipated effect was masked. Doubtless disappointed, Wilson then moved from cosmic-ray research to perfect his cloud chamber and study the causes of lightning.

---

[*] Corresponding author: a.a.watson@leeds.ac.uk

After World War I, the work of Hess was greatly extended by the German physicist, Erich Regener. At a Royal Society Discussion Meeting in 1931 [3] Ernest Rutherford noted that *'there was now reliable data on cosmic ray absorption in water thanks to the far-reaching experiments of Regener'*. Shortly after, Regener pioneered balloon flights in which Geiger counters were carried to great altitudes. He encased his equipment in cellophane to protect against low temperatures: his equipment cooled only to 17º C at the highest altitudes. This was an important technical advance and solved the problems of temperature stability that had blighted similar work by Millikan. In 1935 Regener and his student Pfotzer, succeeded in flying three Geiger counters, arranged to limit the solid angle for detection to within ±20º, to an altitude of 22 km (~40 g cm$^{-2}$) [4], the altitude limit probably being set by the destructive interaction of ozone with the rubber balloons. At an over-burden of ~130 g cm$^{-2}$ they discovered that the coincidence rate maximised, a feature commonly referred to as the 'Pfotzer maximum'. While Pfotzer did follow up and extend this work alone, after Regener had lost his position at the University of Stuttgart because of his political views and because his wife was Jewish, it seems appropriate to refer to this, at the very least, as the Regener-Pfotzer maximum [5].

This discovery, combined with the transition curves reported by Rossi [6], were explained by Bhabha and Heitler [7] who used the newly invented quantum electrodynamics to describe shower phenomena. Schmeiser and Bothe [8] were the first to point out that Rossi's curves implied that there should also be showers in air – naming them '*Luftschauer*' - and showed that particles in showers were separated by up to 40 cm. Clearly Regener's ionisation curve forced the same conclusion. Kolhörster et al [9] extended the separation over which coincidences were found to nearly 80 m. Quite independently, and somewhat serendipitously, Pierre Auger and his colleagues also discovered the air-shower phenomenon. Auger's assistant, Roland Maze, a highly talented electronics expert and a fine physicist, constructed improved Geiger counters and, more importantly, a coincidence circuit with a resolving time of ~5 μs [10]. To determine the resolving power, Auger and Maze used the classical method of measuring the rate of coincidences between two laterally-separated counters, finding that this was much higher than expected by chance. Auger realised how shower theory could explain the phenomenon and very quickly took their equipment to the Pic du Midi (2877 m) and to the Jungfraujoch (3450 m) where they were able to demonstrate coincidences above the chance rate even with counters 300 m apart [11]. With this separation, using two counters, each of 200 cm$^2$, they observed a rate of 1.5 coincidences per hour. Analysis suggested that primaries, assumed to be photons, had energies of ~$10^{15}$ eV [11].

But it was in fact Rossi who observed the air-shower phenomenon first [12]. Working in Eritrea to study the East-West effect, he noted while testing his instruments, '*…that from time to time there arrived upon the equipment very extensive groups of particles (sciami molto estesi di corpuscoli) which produce coincidences between counters even rather distant from each other'*: 'sciami' may be translated as 'swarms'. Rossi was unable to follow up this discovery because of commitments associated with his new position at the University of Padua where he was responsible for the design and equipping of a new physics building. In 1939, because of the racial laws of Mussolini, Rossi and his wife left Italy for the United States. *En route* they spent several months in Manchester where Rossi was hosted by P M S Blackett. Auger, also moving to the USA, was in Manchester at this time. This pre-war interlude offered a remarkable opportunity for three giants of the field to discuss high-energy cosmic rays and was, in retrospect, crucial for high-energy cosmic-ray research in the UK, particularly after the war. Rossi and Auger had become friends while working in the private laboratory of Duc Louis de Broglie in the mid-1930s and Blackett knew both of them well.

Lovell and Wilson (then in their 20s) were directed by Blackett to set up two cloud chambers side-by-side and also with one below the other. According to Lovell [13], the idea for this work probably originated with Auger. Lovell and Wilson observed parallel tracks [14] in the chambers giving striking visual evidence for what had hitherto been inferred indirectly from coincidences between Geiger counters. Blackett was already interested in the shower phenomenon as a result of what he and Occhialini had observed in their counter-controlled cloud chamber in 1933 [15]. Occhialini had studied under Rossi in Florence and brought with him to Cambridge knowledge of Rossi's coincidence circuitry. Cloud chambers had not yet been operated in Italy although one was constructed by Rossi for his Padua laboratory before his forced departure.

After a short time in Manchester, both Rossi and Auger moved to the United States, arriving in time to attend the famous cosmic-ray symposium organised by A H Compton in Chicago in June 1939. In his paper at that meeting [16] Auger speculated that the particles were accelerated to energies of ~$10^{15}$ eV *'along electric fields on a very great extension'*. An even higher energy had been inferred earlier in an analysis made by Janossy and Lovell in 1938 [17] from a shower study using a single cloud-chamber.

Rossi took up the position at Cornell offered to him by Hans Bethe before moving to Los Alamos in 1943 to work on the Manhattan project.

## 3 Work during World War II

Auger, working with M Schein in Chicago, was able to continue studies of air showers particularly through using flotillas of up to 30 balloons [18]. In one flight observations were made at an atmospheric depth of ~120 g cm$^{-2}$ (~15 km). It was concluded that the showers consisted of very narrow and dense bundles of particle: using cascade theory, they deduced that the showers must have been at an early stage of development.

Apart from this experimental work, the only progress in the air-shower field made during WWII was the proposal by Blackett and Lovell to detect air-showers through the reflection of radar from the ionisation trails left by showers. This work was stimulated by Lovell and Wilson noticing traces on radar screens at a station in East Yorkshire on 3 September 1939, two days after the start of World War II. Rather than identifying these as enemy planes, operators claimed that they were due to ionospheric activity. The Blackett and Lovell paper [19] suggesting the use of radar for shower detection is said to have been developed in an air-raid shelter. Shortly after its publication, Eckersley pointed out in a letter to Blackett [20] that the time assumed for the recombination of the electrons in the ionisation trail was one million times too long. It appears that this letter was overlooked, or forgotten, by Blackett during the intensity of his war-effort as shortly after the War ended, Blackett encouraged Lovell to look for showers using the radar-reflection method. The work was carried out at Jodrell Bank where Lovell succeeded in detecting ionisation trails but these were left by meteors. He moved on to play a major role in the development of radio astronomy with the construction of the 250 ft (89 m) dish at Jodrell Bank, now the 'Lovell telescope'. Lovell has acknowledged that had Eckersley's letter not been overlooked, the incentive for the shower search would have vanished and the Jodrell Bank Observatory would not exist today [20].

## 4 Work post WWII

Studies of extensive air-showers were progressed vigorously after the war, particularly in the United States, the United Kingdom, Japan and the USSR. Initially work was done almost entirely with Geiger counters and cloud chambers, the latter used mainly to measure shower directions or, particularly in the USSR, to study what would now be called 'hadronic interactions'. Work using photomultipliers to detect scintillation and Cherenkov light got underway in only the early 1950s.

### 4.1 Work with Geiger Counters in the USSR

The Geiger counter method was exploited particularly effectively for many years in the USSR. A result of outstanding importance was reported by Kulikov and Khristiansen [21] who reported that there was a steepening of the integral spectrum of the number of shower particles at a shower size of ~$10^6$. By modern standards, the effect was not very strongly established, in part because, to claim the effect, it relied on combining the Moscow data with results from another experiment that had covered a higher size regime [22]. Indeed the authors themselves did not regard the irregularity they had observed as totally established because of insufficient statistical accuracy. However the measurement had considerable impact and was verified quickly, and with higher precision, by a number of groups.

The interpretation was unclear but the steepening was an important stimulus for a debate started by B Peters [23] who pointed out that one explanation was that an astrophysical feature at ~$10^{15}$ eV had been observed. Specifically Peters proposed that what was being seen was a consequence of a similar feature in the primary energy spectrum of cosmic rays reflecting a limitation of acceleration within sources or a leakage of particles from the galaxy. These are effects that depend on the rigidity and so on the energy per proton of the primary particle. Thus when protons can no longer be accelerated or they leak out of the galaxy, the flux of cosmic rays falls and the energy spectrum steepens. An alternative proposal was that the feature was due to a characteristic of nuclear interactions with something dramatic happening at around $10^{15}$ eV. The debate, astrophysics *vs* particle physics, was not to be settled for a further 45 years until it was finally resolved by the work of the KASCADE Collaboration in Karlsruhe led by Schatz. Using a complex array of detectors, this group was able to measure the muon content of showers and to deduce the primary energy spectrum as a function of the mass of the incoming particles [24]. They found that the 'knee' in the spectrum of the helium component was at an energy twice that of a similar break in the proton spectrum, thus demonstrating a rigidity effect that could be interpreted as a feature of an acceleration process and/or of leakage of the cosmic rays from the galaxy. The alternative explanation in terms of a change of hadronic interactions that had been advocated in particular by some Soviet scientists was finally ruled out. It is worth noting that Khristiansen provided input to the experimental program developed in Karlsruhe [25].

## 4.2 Work with Geiger Counters in the UK

Work using Geiger counters continued in the UK until the late 1950s. After WWII, while the UK atomic bomb was being developed in the Atomic Energy Research Establishment (AERE) at Harwell, the Director, Sir John Cockcroft, encouraged fundamental research to go on 'outside the wire'. In 1950 B Pontecorvo, who had moved there from Chalk River, Canada, where he had worked with Auger on research related to the Manhattan project, was asked to establish an array of detectors to study air showers. The Harwell work led to some important results on the uniformity of the arrival direction distribution but also to some technical developments that were crucial to future progress.

The first array [26] used only Geiger counters, and so lacked the capability to measure shower directions. Pontecorvo knew of work at MIT, under the guidance of Rossi (see below), to develop a liquid scintillation material that would allow the times at which the shower hit detectors to be measured so enabling the direction of the incoming cosmic-ray to be derived. The scintillator used was a benzene/para-terphenyl mixture. As para-terphenyl was expensive, Pontecorvo asked J V Jelley to investigate the efficiency of the light output as a function of the quantity used. Jelley found that even with no additive he could detect light – Cherenkov radiation - from benzene and from other liquids, including distilled water, as a muon passed through [27]. This discovery led N A Porter to develop a large water-Cherenkov detector [28] subsequently adopted very effectively for shower studies at Haverah Park and at the Pierre Auger Observatory. Porter's detector is compared (figure 1) with one of those used currently at the latter site. Little has changed.

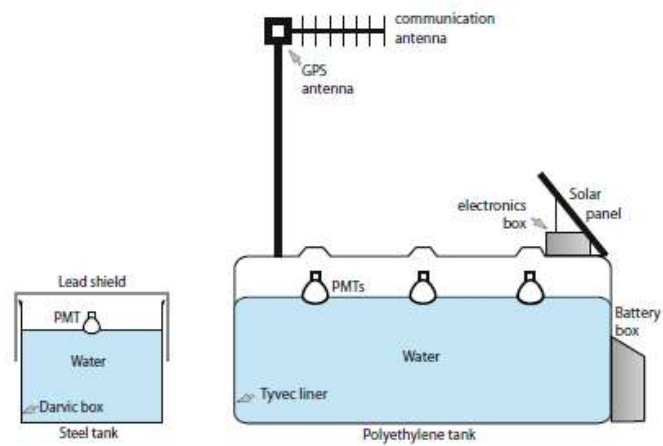

**Fig. 1.** A comparison of the tank (left) developed by Porter [28] and one used at the Pierre Auger Observatory. The detectors have been drawn to the same scale: figure taken from [29].

## 4.3 Air-Cherenkov studies

Blackett was also interested in Cherenkov radiation and was the first to calculate the threshold energies for electrons and muons to produce such light in air [30]. He estimated that light from cosmic rays made up ~$10^{-4}$ of the night-sky background and that signals from air showers might be seen in Cherenkov light using the human eye as a detector. According to Lovell [31], Blackett made an attempt to see these flashes but alas there is no record of his efforts. The idea was taken up by Jelley and W Galbraith, also at Harwell, who adopted a more practical approach. They used a photomultiplier looking downwards at an upward-pointing searchlight mirror and observed signals which they were able to attribute to air-showers [32]. Cherenkov-light work in showers blossomed in places with more congenial climates than that of the UK, notably in the Pamirs and the Crimea. However, Neil Porter developed the technique further after moving to Ireland and one of his students, Trevor Weekes, played a pivotal role in creating the new astronomy of TeV gamma-rays through the detection of a clear signal from the Crab Nebula at the Whipple Observatory in Arizona [33].

The work in the USSR was led by A E Chudakov and G T Zatsepin [34]. It had been realised that most of the energy of the primary particle was dissipated through ionisation of the air and that by determining the 'track-length integral' it would be possible to estimate the primary energy needed to create a shower of a given size using only experimental data. The energy from the particle track-length integral is

$$\varepsilon_c \int_0^\infty N(t)dt,$$

where *N* is the number of particles at a depth *t*. If *t* is expressed in radiation lengths then the energy of the primary is given by multiplying the value of the integral by $\varepsilon_c$, where $\varepsilon_c$ is the critical energy, 84.2 MeV for air. It is assumed that $\varepsilon_c$ is an accurate measure of the average energy lost per radiation length by the charged particles of the shower. The Russians measured the flux of Cherenkov radiation in air showers of N ~$10^5$ to $10^6$, building on the earlier work at the Harwell site.

These data were used independently by Greisen [35] and by Nikolsky [36] to derive an empirical relationship between the electron shower size and primary energy. They were also able to make use of new measurements of the energy and number of muons in showers and of the electromagnetic energy flow. Chudakov et al had shown that in a shower of 1.4 x $10^6$ particles, there were 1.2 x $10^5$ photons from Cherenkov radiation. Assuming that the bulk of the radiation was from electrons of > 50 MeV, Greisen found that the total ionisation loss above the observation level for a shower of this size was 5.2 x $10^{15}$ eV or 3.7 GeV per particle. To get the total energy, 0.2 GeV per electron had to be added for each of the electromagnetic and nuclear particle components with a further 0.4 GeV per particle for the muons and neutrinos increasing the estimate of the primary energy to 6.3 x $10^{15}$ eV.

Before the advent of Monte Carlo calculations, making analytical estimates of fluctuations was very difficult. Nonetheless some insights were made and as often happens these occurred to different people at about the same time. The analyses of Zatsepin [37] and of Cranshaw and Hillas [38] are instructive and reach similar conclusions. By the late 1950s it was recognised from many observations that some properties of showers did not change with energy, altitude or zenith angle in the manner that first considerations led one to expect. For example the attenuation length of shower particles can be deduced from the way in which the rate of showers recorded at a given site varies with barometric pressure. The barometric coefficient is given as $1/\Lambda = - \partial \ln R/\partial t$, where R is the shower rate and t is the atmospheric depth. It can be shown that if $\gamma$ is the slope of the integral spectrum of shower sizes that $\lambda = \gamma\Lambda$, where $\lambda$ is the attenuation length of the shower particles. It was found that $\Lambda$ decreased as the shower size increased and this, combined with the increase of $\gamma$ with shower size, leads to $\lambda$ being approximately constant.

From a number of experiments for showers of size $10^4$ to $10^6$ (around $10^{14}$ to $10^{16}$ eV) it was found that $\lambda$ was nearly constant at ~180 g cm$^{-2}$. This appeared to be consistent with the fact that the shape of the lateral distribution averaged over many showers was also constant for all sizes and altitudes. Only when it became possible to observe showers close to their depth of maximum (~500 g cm$^{-2}$), as at Chacaltaya, did this picture change. However, such constancy is hard to understand if only average development curves are considered. In figure 2 it is evident that the attenuation length, the slope of the shower development curve at the observation depth, would be expected to change with size or with altitude.

Also, it is apparent from consideration of these average curves that the ratio of muons to electrons should change rapidly with altitude or zenith angle, contrary to what is observed. Zatsepin, and Cranshaw and Hillas, came to the same conclusion, namely that the shower cascade from each nucleon coming from a collision is rather short so that at any observation level only the particles in the last cascade, occurring relatively close to the observation level, are observed. The length of the cascade is determined by the energy of $\pi^0$s created in the collisions. Zatsepin memorably described the shower as being like a fir-tree whose branches fluctuate in number and strength. An essential input to this discussion, in addition to the observations on the barometric coefficient and the shape of the lateral distribution, is the assumption that collisions of the proton with a nucleus are elastic so that the out-going nucleon retains a large fraction of the energy.

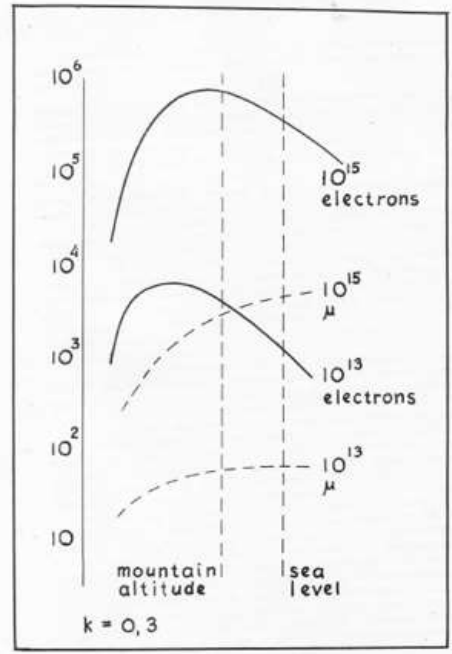

**Fig. 2.** The average longitudinal development of the electron and muon number for primary energies of $10^{13}$ and $10^{15}$ eV assuming an inelasticity of 0.3 [38].

The idea that one cascade dominates what is seen at the observation becomes less tenable when the shower energy increases beyond that needed to produce showers with N ~$10^6$. Thus at energies above $10^{17}$ eV several generations of cascades contribute to what is observed.

Further understanding of fluctuations in showers came as the Monte Carlo technique was developed.

### 4.4 Work of the MIT group

After the war Rossi left Los Alamos to take up a chair at MIT where he established a formidable cosmic-ray group whose '*research program aimed at the study of extensive air-showers, a program which, because of the originality of its conception and the significance of its results, ranks among the foremost accomplishments of the MIT Group*' [39]. This was not an idle boast and most of the methods of analysis of air-shower data from ground arrays that are used currently originate from Rossi's group. A major technical achievement was the development of liquid scintillators using a benzene/para-terphenyl mixture. The fast response of this combination made it possible to measure the arrival direction of showers using electronic methods for the first time [40]. Following a serious fire when one of the scintillators was struck by lightning, a factory to produce solid scintillator blocks was developed [41] and these have played a major role in shower studies ever since.

The work by the MIT group at the Agassiz site was seminal and was preparatory for two ambitious endeavours. One was the BASJE (Bolivian Air Shower Joint Experiment) project mounted in Bolivia, first at El Alto (4200 m) and later at Chacaltaya (5200 m), which was designed to study the 'knee' region of the spectrum with high precision. There was significant input from scientists from Tokyo to these efforts reflecting the strong visitors program that Rossi had established, visits from L Scarsi and M Oda being of particular importance for future ultra-high-energy cosmic-ray (UHECR) activities. The second effort was more exploratory and was led by J Linsley at the Volcano Ranch site (1770 m) in New Mexico. With help from Scarsi, 19 3.26 m$^2$ scintillators were laid out on a hexagonal lattice with 442 m spacing. Linsley then expanded the array to 8 km$^2$, placing the same scintillators 884 m apart. With these arrays Linsley made several ground-breaking findings. With the smaller spacing, and using a 20$^{th}$ scintillator covered by 10 cm of lead, he and Scarsi made the first study of the time distribution of muons and electrons in the shower disc at large distances [42]. With the larger spacing Linsley made the first measurement of the energy spectrum above 1 EeV in which he identified a flattening of the spectrum at a shower size of ~$10^9$ [43] and the first exploration of the distribution of arrival directions above 10 EeV using the 97 most energetic events. In addition he recorded one of the largest showers every detected [44] with an energy claimed to be 100 EeV. Because he chose to run the operation almost single-handedly, and because of severe problems with lightning in New Mexican summers, it was hard to get long on-times. Thus confirmations of Linsley's major findings were not made for many years.

## 4.5 Work in Japan

Before World War II, the Japanese had a vigorous program in both nuclear and cosmic-ray physics. This was led by Nishina and Tomonaga. After the war the two cyclotrons in Nishina's laboratory were thrown into the Bay of Tokyo while those in the universities of Kyoto and Osaka were destroyed [45]: cosmic-ray work continued. Tomonaga led the construction of a cosmic-ray station on Mt Norikura (2700 m) and established the Institute for Nuclear Studies (INS) in Tokyo. Major activity in studying air showers took place there with key contributions coming from Oda and Suga in particular. One of the most notable contributions was the recognition that measurements of $N_\mu$ vs $N_e$ was a way of unravelling the key issue of the mass composition. Again, as with many of the ideas that surfaced in the 1950s and 1960s, the vision was excellent but for its realisation a much larger effort and greater financial investment, such as that in Karlsruhe, was needed.

The work at INS spawned striking efforts leading finally to the construction of an array of detectors covering an area of 100 km$^2$, the AGASA project. This was led first by Kamata and later by Nagano. The detectors used were unshielded scintillators with muons detected using proportional counters covered with concrete. A major conclusion from the work at AGASA was that the energy spectrum extended beyond the steepening predicted by the interactions of protons and nuclei with the microwave background radiation (the GZK effect). It is now known that the interpretation of the data was incorrect, in part because of the systematic uncertainties arising in the estimation of the primary energy using one of the then-favoured models of hadronic interactions.

An important development carried out by Japanese physicists was the analytical calculations of cascade showers produced by photons and electrons carried out by Nishimura and Kamata [46] during and after the war. The starting point for this work was the famous paper by Rossi and Greisen [47] which had reached Japan just before the Japanese joined World War II [48]. The results from the Nishimura-Kamata calculations were of major importance in understanding the longitudinal and lateral development of showers in the days before Monte Carlo calculations and remain of value.

## 4.6 Development of the Fluorescence Technique

In section 4.3 the use of Cherenkov light to help estimate the primary energy of cosmic rays was described. Whilst this was an important breakthrough, the method has been relatively little used at the highest energies ($> 10^{17}$ eV) as the light is emitted in the forward direction so limiting the collecting aperture. This drawback was overcome with the introduction of a second optical method, the fluorescence technique, which has been refined over the last 50 years and is used to make precise measurements of the track-length integral, so enabling the primary energy to be found largely independent of the unknowns of hadronic interactions. The approach relies on the excitation of atmospheric nitrogen by charged particles through the mechanism identical to that responsible for aurorae. The light is emitted isotropically, much of it in the near ultra-violet part of the spectrum, and it has proved possible to detect showers at distances out to ~30 km. The signals are only detectable on clear moonless nights so that accurate and frequent monitoring of atmospheric clarity is essential.

It is unclear who first had the idea of using fluorescence radiation to detect giant air-showers. At the Fifth Inter-American Seminar on Cosmic Rays held in La Paz in 1962, Suga [49] presented preliminary results on fluorescence emission (then called scintillation light) from studies in air using α-particles from polonium and 57 MeV protons to excite the molecules. In the recorded discussion of the meeting Chudakov announced that he had examined the possibility of using the scintillation of air to detect large showers much earlier and had investigated the effect when concerned that it might form a background that would hamper his Cherenkov-light studies. He reported some numbers which were later published [50]. Fluorescence-light studies were never pursued strongly in the USSR.

Greisen was at the La Paz meeting and although he had already started work on the problem at Cornell with his student, Alan Bunner, there are no remarks attributed to him in the Proceedings, nor does he mention the method in his summary report of the meeting [51]. However about a year earlier, in a paper presented at the Kyoto ICRC of 1961, Greisen had commented that several groups throughout the world were studying the possibility of using atmospheric scintillation light to detect giant showers [52]. It is possible that Greisen had picked up the idea from the use at Los Alamos of electromagnetic emission to study atomic bomb detonations at high-altitude, emission that was known as 'Teller radiation'. Whether the Japanese learned of this idea through Oda's visits to MIT through discussions with Rossi (and perhaps Greisen), is unknown and I have found no mention of the fluorescence mechanism in Rossi's writings. Certainly the idea was discussed in Japan at a meeting at Mt Norikura in 1957 with a very recognisable diagram of a detector system shown in the Proceedings [53], figure 3.

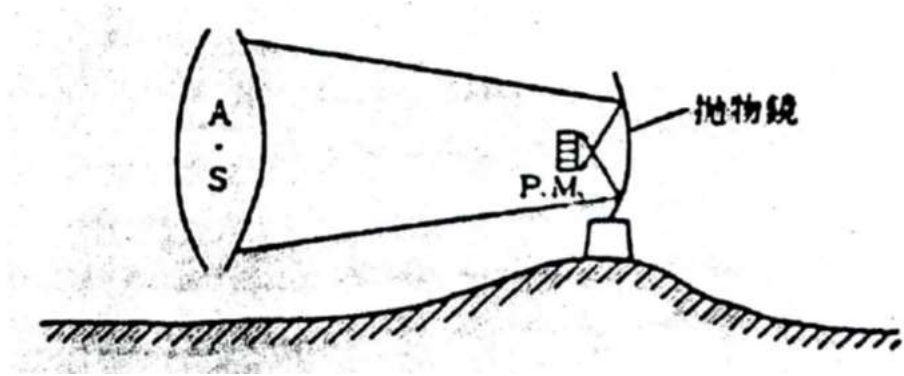

**Fig.3.** Sketch of a fluorescence detector [53].

G Tanahashi spent time working on Greisen's project at Cornell and helped develop the Japanese detector which was used to recorded the first scintillation events in 1969 [54]. The Japanese success is acknowledged in a letter from Greisen to Tanahashi [55] and has recently been re-examined and validated by Dawson [54]. Greisen's own work at Cornell was brought to an end in 1972, handicapped by the poor weather in Ithaca and by the need for more modern electronics. His ideas were quickly picked up by Keuffel's group at the University of Utah leading to the very successful Fly's Eye project [56].

### 4.7 Work in Australia

Work on cosmic rays in Australia was greatly stimulated by the arrival of C B A McCusker at the University of Sydney in 1959. He is best known for his leadership in the construction of an array covering 100 km$^2$ with muon detectors as the elements of the array. The brilliant idea underlying this project was due to McCusker's colleague, M M Winn, who proposed the construction of an array in which each detector operated autonomously [57]. A timing signal was projected across the area so that the instant at which a trigger occurred could be recorded locally: this time, along with signal amplitude, was registered on magnetic tape. The concept is exactly that used at the Auger Observatory and by the Telescope Array Collaboration except that modern technology, including the availability of GPS and of cellular telephone developments, has made it much easier to implement Winn's concept than it was in the 1960s and 1970s. At this time, coincidences between stations that define events could only be discovered off-line many days later through inspection of the magnetic tapes from the detectors at Sydney University.

### 4.8 Monte Carlo Calculations

Advances in the availability of high-performance computers from the mid-1960s led to new insights. Leaders in this effort were A M Hillas, P K F Grieder and J-N Capdevielle. Programs developed by the latter two played an important role to the creation of the CORSIKA code by the Karlsruhe Group, now used world-wide to help interpret shower data [58].

Hillas introduced several innovations to Monte Carlo methods (e.g. the splitting algorithm [59]) but most notably he pointed out that for any particular spacing of detectors in an array, there was a distance at which the signal size could be measured with small uncertainty. This quantity could be used to characterise the size of a shower without having detailed knowledge of the lateral distribution of the signals [60]. The idea was first introduced in the context of analysis of data from Haverah Park where it was realised at an early stage that converting the observations made in water-Cherenkov detectors to the number of electrons in a shower (the shower size used historically because of early work with Geiger counters) was quite impractical. For the first stage of the Haverah Park project the signal at 500 m from the shower axis was adopted with 600 m being chosen for the final 12 km$^2$ array. Although the idea was slow to catch on and was met with considerable opposition, it is now the size parameter of choice at a number of installations: S(125) is used for IceTop, S(800) at the Telescope Array and S(1000) at the Pierre Auger Observatory.

A good summary of the status of Cascade Simulations as at 1981 was prepared by Linsley and Hillas [61] who attempted, with limited success, to get simulators to run their codes with common input parameters for multiplicity, mean free path, inelasticity etc..

### 4.9 Radio technique 1965 – 1975

In 1965 radio emission at 44 MHz was detected in coincidence with extensive air-showers by a team led by Jelley that included Porter and Weekes [62]. They had been searching for the Cherenkov-like emission predicted by Askaryan [63] to arise because of the negative charge-excess expected in extensive air-showers. Signals were detected using an array of 72 full-wave dipole antennas operated in coincidence with three Geiger counter trays. This work stimulated studies of showers using radio across many countries most notably in the USSR at Moscow State University Array (Khristiansen) and at Haverah Park where the work was led by H R Allan. The emission mechanism was found to be dominated by radiation from the dipole established through geomagnetic separation of the electrons and positrons in the showers [64]. The lateral distribution of the radio emission is steep, like that of the electrons, so thwarting early ambitions that a technique had been found to allow large arrays to be built economically. However sensitivity of the amplitude of the signal to the depth of shower maximum was recognised and the prospect of determining the mass composition was seriously explored until it was discovered that variations in the geoelectric field could produce large changes in the signal sizes detected [65]. This led to the technique being largely abandoned in the mid-1970s.

Some 30 years later, with improved technology to record the signals and with the capability of measuring the state of the geoelectric field easily, the radio technique has been revived and is showing great promise for studies of charged cosmic rays [66] and for the detection of neutrinos in ice [67].

### 4.10 Detecting fluorescence radiation from Space

All of the tools for studying air showers described above were proposed over 50 years ago. The most recent innovation came from Linsley who pushed the concept of detecting air showers from space from 1979 [68]. Who first thought of this approach is unclear with Linsley recalling the idea being put to him at an astrophysics planning meeting probably in the late 1970s. But the idea may have been in the air earlier as Bunner has commented [69]: *'I also recall having the idea, about 1961-62, of using a satellite to look down on the night-time Earth to extend the effective area. I'm sure that others independently thought of that idea too. We also talked about whether the atmospheres of other planets might work'*.

Several projects are underway with the goal of achieving Linsley's 1979 vision (see section 7.3 below).

### 4.11 The growth and impact of large collaborations

Until the late 1980s the collaborations studying extensive air-showers were quite small. Typically they comprised only a few people with papers rarely signed by as many as ten authors. This changed significantly with the development of the KASCADE detector [70] in Karlsruhe that was built with the purpose of making precise measurements of the mass composition in the region of the spectrum knee around 1 PeV and the CASA array [71] that was constructed in the Dugway desert by a team led by J W Cronin to study *inter alia* gamma-ray emission at ~1 PeV from Cygnus X-3. Typically 20 – 40 authors signed papers from these collaborations. The leaders of both projects had been accustomed to the generous funding available for nuclear and particle physics and brought a wealth of experience to these bold endeavours.

The size of collaborations has continued to grow with the IceTop, Telescope Array and Auger Collaborations regularly producing papers with author lists ranging from 100 to over 400. The days of the iconoclastic cosmic-ray worker labouring alone have long gone, though of course leadership of such large endeavours requires a different portfolio of skills. No one, however, can be single-handedly responsible for the conception, implementation and analysis efforts that these giant projects demand.

The growth of the large collaborations necessary to implement the ideas of the early visionaries has brought us a much more precise understanding of features of the energy spectrum, the arrival direction distributions and the mass composition from below $10^{15}$ eV to nearly $10^{20}$ eV.

### 5. Role of theory and phenomenology in guiding experiment

In many fields of science experimental discoveries march hand-in-hand with theoretical developments. This is not the case for studies of high-energy cosmic rays. The knee in the spectrum was discovered by observation as noted above. Various theories to explain it have been explored over the years but there was no prediction made of what was observed. At the very highest energies, the discovery of the 2.7 K background radiation in 1965 led Greisen [72] and Zatsepin and Kuzmin [73] to predict an abrupt steepening of the energy spectrum. Although searching for this effect was certainly a goal of arrays such as AGASA, a further motivation for building larger and more complex projects was the belief that eventually the particles would detected with sufficient rigidity to cut through the intervening magnetic fields with deflections sufficiently small for UHECR astronomy to begin.

No phenomenology has had the power to predict the flux of cosmic rays expected at any energy: the sources of production and the acceleration processes are too little understood and thus there has been no guidance as to how large a detector should be constructed. When promoting the Auger Observatory, one of the hardest questions to deal with was '*why do you want to make it so big?*'

## 6 The Present

A major objective of the UHECR series of meetings is to discuss the progress that has been made by the Working Groups established to attempt to reach agreed positions about work on the mass composition, the energy spectrum and the arrival directions distributions. Results of these efforts are described in accompanying papers by A Yushkov, D Ivanov and J Biteau respectively. I want to make a few remarks about the outcomes people should look for in the reports from two of the Groups, those for the mass composition and energy spectrum.

### 6.1 Mass Composition as a function of energy

Progress has been made through the efforts of the Group on mass composition on the situation in the energy range $18.2 < \log E \text{ (eV)} < 19.0$. This was reported in a joint paper presented during the 2017 ICRC in Busan [74]. The agreed conclusion is that the TA $X_{max}$ distributions in this energy range are as compatible with a pure proton composition as they are with the mixed composition reported by the Auger Collaboration. I would urge the Community to abide by the exhortations made by V de Souza in his Busan talk. In evaluating the conclusions, it must be recognised that the detailed comparisons demand use of the simulations of the performance of the two detector systems and detailed cross-checks of the analyses methods of both collaborations. Phenomenologists should refer to this joint analysis, bearing in mind that it refers to a restricted range in energy. There is no controversy between the Collaborations over this conclusion – although, admittedly, it is not as strong as one might wish.

The situation at energies above 10 EeV is less clear. The Auger Collaboration has claimed for some time that the mean mass increases as a function of energy [75]. This conclusion is based both on the manner in which $X_{max}$ and $\sigma(X_{max})$ change with energy and on the distributions of the $X_{max}$ measurements in narrow energy intervals: the data from the Telescope Array are as yet too sparse to allow the kind of detailed analysis presented in [75].

Is there a simpler way to get an answer to the important question as to whether or not the mass does change? I would argue that a comparison of elongation rates is a straightforward and model-independent way to proceed. Using the data published recently by TA [76] (specifically the numbers listed in table 4 that accompanies figure 14 of this paper), I find that the overall fit, in the range $18.2 < \log (E/\text{eV}) < 19.6$, gives an elongation rate of $(55 \pm 3)$ g cm$^{-2}$ with a reduced $\chi^2$ of 3.04 (for 9 degrees of freedom, $p < 10^{-3}$). By contrast the values quoted in the TA paper are $(57 \pm 5)$ g cm$^{-2}$ with a reduced $\chi^2$ of 1.19 ($p \sim 0.3$). One does not need to calculate $\chi^2$ to see that a fit to single-line description of the data will be poor. The $X_{max}$ data are an unsatisfactory fit to a straight line (figure 4) with 6 (nearly 8) of the 11 points lying more that 1 sigma from the line whereas 3.7 would be expected – and some points are quite far from the line. The reason for the difference between the two $\chi^2$ evaluations was not resolved during the meeting and the issue is not discussed in the paper by W Hanlon prepared for these Proceedings [77].

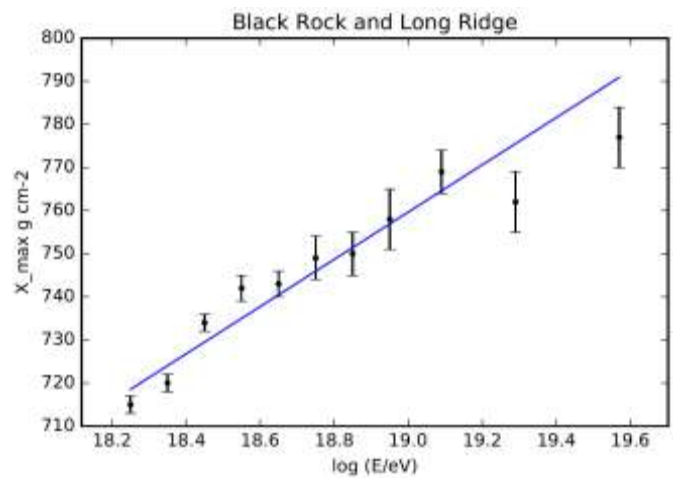

**Fig. 4.** Data from table 4 of [76] plotted for this paper.

The variations of $X_{max}$ with energy reported by each Collaboration are shown in figure 5. The same flattening of the elongation rate, long noted by the Auger Collaboration, is evident in the plot of the TA data [76]. Above log E (eV) = 18.54, the elongation rate is (37 ± 6 g cm$^{-2}$), significantly different from the elongation rate of (95 ± 14) g cm$^{-2}$ at lower energies. This is surely evidence for the mean mass rising with energy (or for a change in the hadronic physics). For the reasons discussed in the Working Group report, a point-by-point comparison of absolute values of $X_{max}$ cannot be made but this stricture does not apply to evaluations of the Elongation Rates.

The measurements shown in figure 5, made using fluorescence detectors, are presently limited to energies below 40 EeV because of the small number of events. To extend the elongation rate to higher energies it is necessary to use the detectors of the surface arrays that are operated with nearly 100% on-time. At the Auger Observatory a method has been devised that extends measurements to nearly 100 EeV using the risetimes of the signals in the water-Cherenkov detectors and it has been found that the elongation rate is a smooth continuation of that measured below 40 EeV (figure 6), implying that the mean mass is continuing to increase as the energy rises [79].

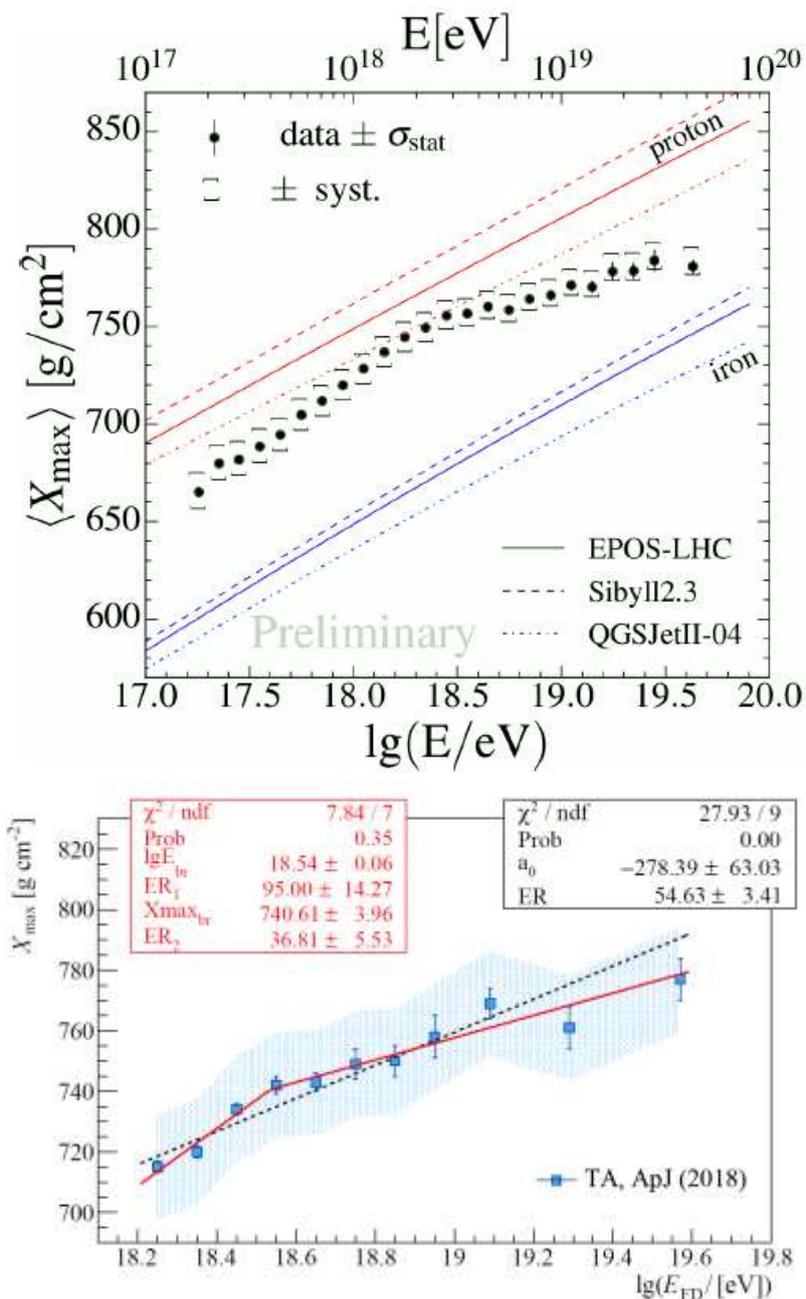

**Fig. 5.** Elongation rates reported by the Auger Collaboration [78] (upper figure) and extracted from the data reported by the Telescope Array Collaboration in [76] (lower figure).

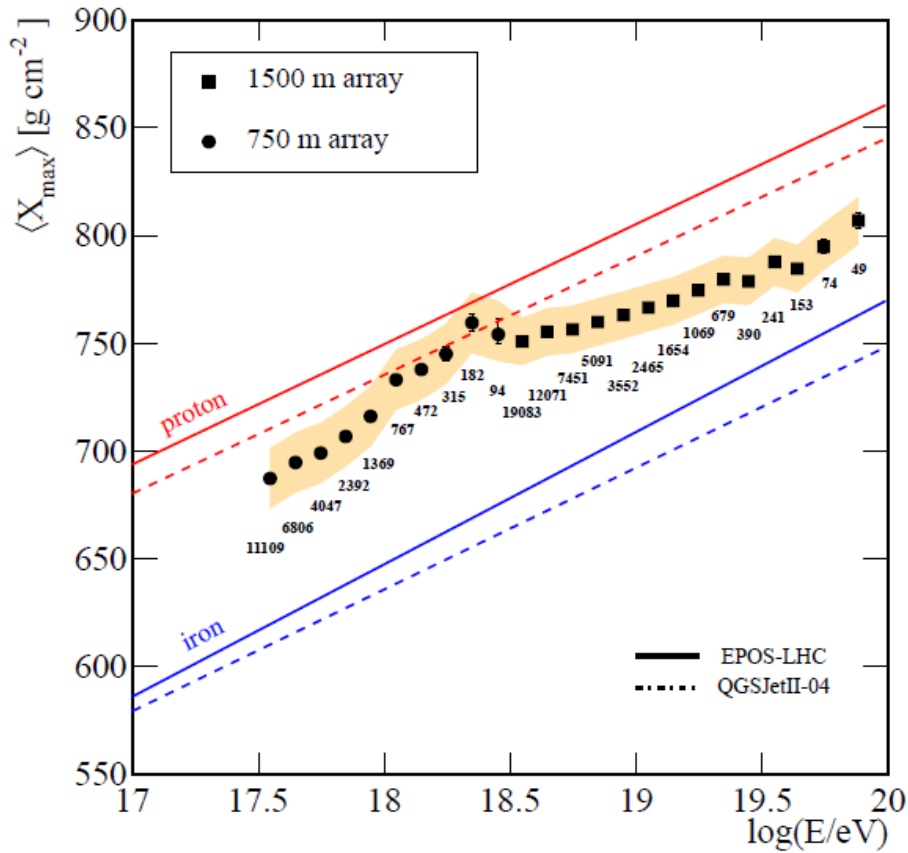

**Fig.6.** Estimates of $X_{max}$ deduced from the risetime of the signals in the water-Cherenkov detectors of the Auger Observatory [79].

**6.2 The Energy Spectrum**

It is generally agreed that the spectrum flattens at about 5 EeV (the ankle) and then steepens abruptly above about 40 EeV. In the ankle region agreement between the measurements made by TA and the Auger Collaborations can be improved by shifting the energy scales by ±5%, the Auger estimate being increased and the TA estimate reduced. These are small changes and are well-within the systematic uncertainties reported by each group.

However the situation at higher energies is much less satisfactory. In my concluding remarks at the 2012 incarnation of these UHECR meetings, I wrote [80] *'I believe that it is highly desirable for the Working Group on the Energy Spectrum to move next to a comparison of spectra from the same part of the sky. There are useful overlaps with TA and Auger and between TA and Yakutsk. I believe that there is much to be learned by making such comparisons and this would at the same time explore whether or not fluxes seen in the Northern and Southern hemispheres are different as they might be if, for example, a local source such as Cen A is dominant in the data set recorded at the Auger Observatory.'* These remarks were made before the report of a 'Hot Spot' by the TA group [81] which serves only to heighten the importance of this point.

Progress in making this cross-check has been rather slow. In a paper posted on the archive server in January 2018 [82], submitted to Astroparticle Physics (and still with referees at the time of writing), the TA group claim that their spectrum at high energies has a greater flux than that reported by the Auger Collaboration with the excess being at the high declinations inaccessible to the Auger detector. In [82] they address the measurements in the common declination band but their analysis can be criticised. First the energy estimate from the Auger Collaboration is increased by 11%, contrary to the ± 5% shifts agreed in the Working Group. They then find what they call a 'break energy' by fitting the spectra from the Collaborations with two power laws. The break energies agree at the 0.5 sigma level and is used to justify a claim *'(at the 3.5 sigma level) for a change in the energy spectrum of ultra-high energy cosmic rays in the northern hemisphere of the sky'*.

Such a strong claim can only be believed if the flux discrepancy in the common declination band is thoroughly understood: I do not believe that it is. It is evident from the data plotted in figure 3 of [82] for the common declination band, -15.7º < δ < 24.8º, that there are statistically significant differences between the slopes: the TA spectrum is harder than the Auger spectrum. Specifically the slopes are (-2.94 ± 0.04) and (-2.71 ± 0.04) for the Auger and TA data respectively. Perhaps both measurements are wrong: both cannot be right. Note also that the conclusion from the elongation rate discussion in 6.1, implies that the TA energy estimates need to be revised in two ways to accommodate a non-protonic mass. Firstly the adjustment to energies measured with the fluorescence method that must be made for muons and neutrinos carried into the earth will be larger than currently adopted, and secondly the primary energy needed to produce a particular S(800) will be greater. This latter correction is also dependent on the hadronic model.

## 7 The Future

### 7.1 Mass Composition question

The curent need to use hadronic physics to determine the mass makes it hard to see how the question of composition can finally be resolved. Anyone strongly wedded to the proton-hypothesis can always argue that the hadronic physics is different at centre-of-mass energies > 10 times that achieved at the LHC. I see two lines of attack that may give hope of resolving this dilemma.

Zatsepin, and Zatsepin and Gerasimova, [83] pointed out in the 1950s that photodisintegration in the photon field around the sun might lead to widely separated showers at ground level. This effect has been re-examined [84] using current information about the inter-planetary magnetic field and it is evident that an array much larger than the present Auger area will be required: observing this phenomenon may form part of the science case for future instruments. However this effect has the potential to give us mass data only around 1 EeV. At higher energies we may hope, with hugely increased event numbers, to identify a small number of sources unambiguously, and then use the galactic magnetic field as a magnetic spectrometer. Such approaches require new instruments.

### 7.2 Upgrades of existing ground arrays

Upgrades of both the Auger Observatory and the Telescope Array are underway and details are given in these Proceedings by A Castellina (Auger) and E Kido (Telescope Array).

The Auger Collaboration are adding 4 m$^2$ of thin scintillator above the water-Cherenkov detectors along with faster electronics and an extension to the dynamic range through the addition of a small photomultiplier to each detector. The major physics goals are the identification of light primaries in the decade above 10 EeV on an event-by-event basis which will allow composition-sensitive searches of arrival direction patterns. Additionally unexpected changes in hadronic interactions will be sought.

An important development for the Auger Observatory will be the addition of a radio antenna at each of the 1600 detectors on the 1500 m hexagonal grid. This important step is described in these Proceedings by J Hörandel. A particularly exciting target is the independent measurement of the electromagnetic component of inclined showers using the radio antennas simultaneously with a high-precision measurement of the muon component. The photons and electrons of the electromagentic component are 'ranged out' and do not reach the ground whilst the muons survive. These measurements are expected to lead to further insights about the mass composition at high energies.

At the Telescope Array work is underway to increase the area by a factor of 4 (TAx4) by adding 500 more scintillators. This enlarged area will be similar to that of the Auger Observatory. The spacing of 2.08 km means that the array will be 95% effective above 57 EeV with an angular resolution of 2.2º. The main physics goals are a better understanding of the energy spectrum and higher statistics in the TA Hot Spot region.

### 7.3 Prospects from Space

Linsley's ideas (see section 4.10) were first developed by a team lead by Scarsi and himself through a proposal to ESA under the name EUSO, the Extreme Universe Space Observatory. Despite a successful Phase A study completed in 2014, ESA decided not to proceed with the mission to fly a fluorescence detector because of financial constraints. The program was then taken up by the Japanese space agency (Japanese Experiment Module-EUSO or JEM-EUSO) with support from over 90 institutions from 16 countries. At present several developments under the JEM-EUSO umbrella, with JEM now standing for Joint Experimental Missions, are underway. These include ground-based studies undertaken at the Telescope Array (M Bertaina in these Proceedings) and the successful operation in space of the TUS instrument, a pathfinder for KYLPVE-EUSO, a Russian-led effort. The important TUS results, from the first fluorescence detector operated in space, are described in these Proceedings by P Klimov.

American-led efforts which, in addition to the work at the Telescope Array, have so far involved two balloon flights, one using the NSF long-duration facility, are targeted at the POEMMA mission which is envisaged to have fluorescence detectors on two spacecraft to give stereo images of showers from space. The status of this mission is described in detail by J Krizmanic in these Proceedings. Scientists from 16 institutions are seeking support from NASA to develop a proposal that will be considered during the 2020 Astronomy and Astrophysics Decadal Review in the USA.

### 7.3 Longer term prospects

The results from space experiments, assuming that they are as exciting as anticipated, will need to be followed up by a massive effort on the ground. As noted before [80], I see the situation as being analogous to the LEP machine and the associated detectors that were built to study the W and Z properties in exquisite detail. KLYPVE-EUSO and POEMMA might be thought of being analogous to the SPS and one can envisage a future in which space observations are followed up using one or more dedicated ground arrays of enormous size covering more than 30,000 km$^2$ and of a hybrid nature. The scale of such a project can surely only be achieved through an international effort perhaps to be realised in 8 to 10 years, but it is not too early to discuss plans, form collaborations and carry out R&D.

To give full sky coverage from the ground one requires two observatories at latitudes similar to those selected by the Auger and TA Collaborations, as pointed out many years ago in the Design Report for the Auger Observatory [85]. It may be possible to link with one of the other massive projects underway or planned (e.g. SKA with sites in South Africa and Australia or the GRAND site in China) to avoid duplication of infrastructure and negotiations with land-owners. I can suggest no new techniques that can be added to the battery of methods that we already have so that it is going to be necessary to make present instrumentation more cheaply.

**Acknowledgements:** I am most grateful to the Local Organising Committee for the invitation to give the talk that led to this paper and for their financial support. In its preparation, I have drawn quite heavily on research that Karl-Heinz Kampert and I carried out for our article on the history of extensive air-showers [29] and remain very grateful for his support. I am also indebted to Hiro Sagawa for tracking down details for reference 53 and thanks are also due to Alexey Yushkov who created the lower plot of figure 5. I have benefitted enormously from interactions with my many friends and colleagues within the Pierre Auger Collaboration over the years and thank some of them for input that has led to the conclusions given towards the end of this paper. However, I cannot stress too strongly that the opinions expressed in section 6 may not represent the views of all of the Collaboration.

## References

1. V L Fitch, Rev Mod Phys **71** S25 (1999)
2. C T R Wilson, Proc Roy Soc A **68** 151 (1901)
3. Royal Society Discussion Meeting on 'Ultra Penetrating Rays', Proc Roy Soc A London **132** 331 (1931)
4. E Regener and G Pfotzer, Nature **136** 718 (1935)
5. P Carlson and A A Watson, History of Geo- and Space Sciences **5** 175 (2014)
6. B Rossi, Zeitschrift für Physik **82** 151 (1933)
7. H J Bhabha and W Heitler, Proc Roy Soc A **159** 432 (1937)
8. K Schmeiser and W Bothe, Ann Phys **424** 161 (1938)
9. W Kolhörster, I Matthes and E Weber, Naturwissenschaften **26** 576 (1938)
10. R Maze, J Physique et le Radium **9** 162 (1938)
11. P Auger, R Maze and Robley, Compte Rendue **208** 1641 (1939)
12. B Rossi, Supplemento a la Ricerca Scientifica **1** 579 (1934)
13. A C B Lovell to A A Watson (private communication) 24 January 2008

14. A C B Lovell and J G Wilson, Nature **144** 863 (1939)
15. P M S Blackett and G Occhialini, Nature **130** 363 (1932)
16. P Auger et al., Rev Mod Phys **11** 288 (1939)
17. L Janossy and A C B Lovell, Nature **142** 716 (1938)
18. P Auger, A Rogozinsk and M Schein, Phys Rev **67** 62 (1945)
19. P M S Blackett and A C B Lovell, Proc Roy Soc Lond A **177** 183 (1941)
20. T L Eckersley to P M S Blackett, 12 March 1941. For details see A C B Lovell Notes Rec Roy Soc Lond. **47** 119 (1993)
21. G V Kulikov and G B Khristiansen, Soviet JETP **35** 441 (1959)
22. L K Eidus et al., Soviet JETP **22** 440 (1952)
23. B Peters, Nuovo Cimento **22** 800 (1961)
24. T Antoni et al., Astroparticle Phys **24** 1 (2005)
25. Notes from G B Khristiansen to G Schatz, late 1980s (unpublished)
26. T E Cranshaw and W Galbraith, Phil Mag **45** 1109 (1954)
27. J V Jelley, Proc Phys Soc A **64** 82 (1951)
28. N A Porter et al., Phil Mag **3** 826 (1958)
29. K-H Kampert and A A Watson, European Physical Journal H **37** 359 (2012); arXiv:1207.4827
30. P M S Blackett, Physical Society of London Gassiot Committee Report 34 (1948)
31. A C B Lovell, Biographical Memoirs of Royal Society **21** 1 (1975)
32. W Galbraith and J V Jelley, Nature **171** 349 (1953)
33. T C Weekes et al. ApJ **342** 379 (1989)
34. A E Chudakov et al., Proc Moscow ICRC II 50 (1960)
35. K Greisen, Progress in Cosmic Ray Physics **III** 3 (1960), North Holland
36. S I Nikolsky, Proc 5[th] Inter-American Seminar on Cosmic Rays, La Paz, **2** XLVIII (1962)
37. G T Zatsepin, Proc ICRC Moscow **II** 192 (1960)
38. T E Cranshaw and A M Hillas, Proc ICRC Moscow Vol **II** 210 (1960)
39. B Rossi, *'Moments in the Life of a Scientist'* (1990), Cambridge University Press
40. P Bassi, G Clark and B Rossi, Phys Rev **92** 441 (1953)
41. G W Clark, F Scherb and W B Smith, Rev Sci Inst **28** 433 (1959)
42. J Linsley and L Scarsi, Phys Rev **128** 2384 (1962)
43. J Linsley. Proc 8[th] ICRC Jaipur **4** 77 (1963)
44. J Linsley, Phys Rev Letters **10** 146 (1963)
45. J Nishimura, AAPS Bulletin **17** Feb (2007)
46. K Kamata and J Nishimura, Suppl Prog Theor Physics **9** 93 (1958)
47. B Rossi and K Greisen, Rev Mod Phys **13** 240 (1941)
48. J Nishimura to A A Watson, 5 July 2015 (private communication)
49. K Suga, Proc 5[th] Inter-American Seminar on Cosmic Rays, La Paz, **2** XLIX-1(1962)
50. V A Belyaev and A E Chudakov, Bull Acad Science of the USSR **30** 1700 (1966)
51. K Greisen, Physics Today April 58 – 60 (1963)
52. J Delavaille, F Kendziorski and K Greisen, Proc 7[th] ICRC (Kyoto) J Phys Soc Japan **17** Supple A-III 76 (1962)
53. K Murayama (editor), Norikura Symposium Report, Cosmic-Ray Research **3**, No.5 449 (1958)
54. T Hara et al., Acta Phys Acad Sci Hungaricae **29** 369 (1970) *and* B Dawson, arXiv:1112.5686
55. K Greisen, letter to G Tanahashi, 29 Sept 1969
56. R M Baltrusaitis et al. Nucl Inst Meth **A 240** 410 (1985)
57. C B A McCusker and M M Winn, Il Nuovo Cimento **28** 175 (1963) and Proc 8[th] ICRC Jaipur **4** 306 1963
58. D Heck et al., Forschungszentrum Karlsruhe Report FZKA 6019 1988
59. A M Hillas, Proc 17[th] ICRC Paris **8** 193 (1981)
60. A M Hillas, Acta Phys Acad Sci Hungaricae 29 355 1970
61. J Linsley and A M Hillas (editors), Proc Paris Workshop on Cascade Simulations Paris 1981
62. J V Jelley et al., Nature 205 327 (1965)
63. G A Askaryan, Sov Phys JETP 14 441 1962
64. H R Allan, Progress in Elementary Particle and Cosmic Ray Physics 10 169 1971 North Holland
65. N Mandolesi, G Morigi and G Palumbo, J Atm Terr Phys **36** 1431 (1974)
66. Pierre Auger Collaboration, Proc 34[th] ICRC The Hague Proc Sci 420 (2015): arXiv:1509.03732
    S Buitink et al. Nature **531** 70 (2016)
67. P Allison et al., Astropart. Phys **35** 457 (2012)
68. J Linsley, Proposal to the Field Astronomy Survey Committee 1979, and La Jolla ICRC **3** 438 (1985)
69. A Bunner to A A Watson (private communication), July 2010
70. T Antoni et al., NIMA **513** 490 (2003)
71. A Borione et al., NIMA **346** 323 (1994)
72. K Greisen, Phys Rev Letters **16** 748 (1966)
73. G T Zatsepin and V A Kuzmin, Soviet JETP Letters **4** 78 (1966)

74. V de Souza, Pierre Auger and Telescope Array Collaborations, Proc 35$^{th}$ ICRC Busan 2017; arXiv:1801.01018
75. A Aab et al., Pierre Auger Collaboration, Phys Rev D **90** 122006 (2014)
76. R U Abbasi et al., Telescope Array Collaboration, Ap J **858** 76 (2018)
77. W Hanlon, Telescope Array Collaboration: Proceedings of UHECR2018, arXiv:1812.05688
78. J Bellido for the Pierre Auger Collaboration, 35$^{th}$ ICRC Busan 2017; arXiv:1708.06592v1
79. A Aab et al., Phys Rev D **96** 122003 (2017)
80. A A Watson, Proceedings of UHECR2012 EPJ Web of Conference **53** 02003 (2013)
81. R U Abbasi et al, Telescope Array Collaboration, Astrophys J 790 L21 2014: arXiv:1404.5890
82. R U Abbasi et al, Telescope Array Collaboration, arXiv:1801.07820
83. G T Zatsepin, Dokl Acad Nauk SSSR **80** 577 (1951)
    N M Gerasimova and G T Zatsepin, Soviet Physics JETP **11** 899 (1960)
84. G A Medina-Tanco and A A Watson, Astroparticle Phys **10** 157 (1999); astro-ph/9808033
    L N Epele, S Mollerach and E Roulet, J High Energy Physics **3** 017 (1999); astro-ph/981230
85. Auger Design Report, Fermi National Accelerator Laboratory October 1995